\documentclass[a4paper,11pt]{article}

\usepackage{amsfonts}
\usepackage{amsmath}
\usepackage{amssymb}
\usepackage{authblk}
\usepackage{cancel}
\usepackage[font={footnotesize,it}]{caption}
\usepackage{cite}
\usepackage{color}
\usepackage{comment}
\usepackage{default}
\usepackage{enumitem}
\usepackage{epsfig}
\usepackage{float}
\usepackage[margin=2.5cm]{geometry}
\usepackage{graphicx}
\usepackage{hyperref}
\usepackage{cleveref} % `cleveref' cannot be called before `hyperref'
\usepackage[utf8]{inputenc}
\usepackage{mathtools}
\usepackage{ragged2e}
\usepackage{subcaption}
\usepackage{tikz}
\hypersetup{
citecolor=red,
colorlinks=true,
filecolor=red,
linkcolor=blue,
linktocpage=true,
urlcolor=blue
}

{\makeatletter \g@addto@macro\bfseries{\boldmath} \makeatother}

\usetikzlibrary{shapes}

\begin{document}

\title{Holographic entanglement negativity for disjoint intervals in $AdS_3/CFT_2$}

\author[1]{Vinay Malvimat\thanks{\noindent E-mail:~ {\tt vinaymmp@gmail.com}}}
\author[2]{Sayid Mondal\thanks{\noindent E-mail:~ {\tt sayidphy@iitk.ac.in}}}
\author[2]{Boudhayan Paul\thanks{\noindent E-mail:~ {\tt paul@iitk.ac.in}}}
\author[2]{Gautam Sengupta\thanks{\noindent E-mail:~ {\tt sengupta@iitk.ac.in}}}

\affil[1]{
Indian Institute of Science Education and Research\\

Homi Bhabha Rd, Pashan, Pune 411 008, India
\bigskip
}

\affil[2]{
Department of Physics\\

Indian Institute of Technology\\ 

Kanpur 208 016, India
}

\date{}

\maketitle

\thispagestyle{empty}

\begin{abstract}

\noindent

\justify

We advance a holographic construction for the entanglement negativity of bipartite mixed state configurations of two disjoint intervals in $(1+1)$ dimensional conformal field theories ($CFT_{1+1}$) through the $AdS_3/CFT_2$ correspondence. Our construction constitutes the large central charge analysis of the entanglement negativity for mixed states under consideration and involves a specific algebraic sum of bulk space like geodesics anchored on appropriate intervals in the dual $CFT_{1+1}$. The construction is utilized to compute the holographic entanglement negativity for such mixed states in $CFT_{1+1}$s dual to bulk pure $AdS_3$ geometries and BTZ black holes respectively. Our analysis exactly reproduces the universal features of corresponding replica technique results in the large central charge limit which serves as a consistency check.

\end{abstract}

\clearpage

\tableofcontents

\clearpage

\section{Introduction}
\label{sec_intro}

Quantum entanglement has attracted intense focus recently in diverse disciplines from condensed matter physics to issues of quantum gravity \cite{VanRaamsdonk:2009ar,VanRaamsdonk:2010pw,Swingle:2009bg,Maldacena:2013xja,Hartman:2013qma}. The entanglement for bipartite pure states may be characterized by the entanglement entropy which is defined as the von Neumann entropy of the reduced density matrix for the subsystem under consideration. However entanglement entropy fails to be a viable measure for the characterization of mixed state entanglement as it incorporates correlations irrelevant to the specific bipartite system in question. This significant issue in quantum information theory was addressed by Vidal and Werner in \cite{Vidal:2002zz}, where they introduced a computable measure termed entanglement negativity which characterized the upper bound on the distillable entanglement for the bipartite mixed state.\footnote{Distillable entanglement characterizes the amount of pure entanglement that can be extracted from the state in question using only LOCC.} The entanglement negativity was defined as the logarithm of the trace norm of the partially transposed density matrix with respect to one of the subsystems of a bipartite system. It was shown by Plenio in \cite{Plenio:2005cwa} that the entanglement negativity was not convex but was an entanglement monotone under local operations and classical communication (LOCC).

In \cite{Calabrese:2004eu,Calabrese:2009qy,Calabrese:2009ez,Calabrese:2010he} the authors advanced a comprehensive procedure to compute the entanglement entropy in $(1+1)$ dimensional conformal field theories ($CFT_{1+1}$) employing a replica technique. For configurations involving multiple disjoint intervals the entanglement entropy computed through the replica technique receives non universal contributions which depend on the full operator content of the $CFT_{1+1}$. It was later shown in \cite{Hartman:2013mia,Headrick:2010zt} that these non universal contributions were sub leading in the large central charge limit. Subsequently a variant of the above replica technique could be utilized to compute the entanglement negativity of various bipartite pure and mixed state configurations in a $CFT_{1+1}$ \cite{Calabrese:2012ew,Calabrese:2012nk,Calabrese:2014yza}. Interestingly the entanglement negativity for a bipartite pure state was given by the R\'{e}nyi entropy of order half in conformity with quantum information theory. Following this, in \cite{Kulaxizi:2014nma} the large central charge limit of the entanglement negativity for a mixed state configuration of two disjoint intervals was investigated. Interestingly in this case the entanglement negativity is non universal in general except when the two intervals are in proximity where a universal contribution may be extracted in the large central charge limit \cite{Kulaxizi:2014nma}. Remarkably through a monodromy analysis it could be numerically demonstrated that the entanglement negativity exhibited a phase transition \cite{Kulaxizi:2014nma,Dong:2018esp}.

In the context of the $AdS/CFT$ correspondence Ryu and Takayanagi (RT) \cite{Ryu:2006bv,Ryu:2006ef} advanced a holographic conjecture to describe the universal part of the entanglement entropy of a subsystem in a dual $CFT_d$. This was given by the area of the co dimension two static minimal surface in the bulk $AdS_{d+1}$ geometry, homologous to the subsystem. This development attracted intense interest in obtaining the holographic entanglement entropy of bipartite systems described by dual $CFT_d$s (for a detailed review see \cite{Ryu:2006ef,Nishioka:2009un,Rangamani:2016dms,Nishioka:2018khk} and references therein). A covariant generalization of the Ryu-Takayanagi conjecture was subsequently advanced in \cite{Hubeny:2007xt} by Hubeny, Rangamani and Takayanagi (HRT). A proof of the RT conjecture was established from the bulk perspective initially in the context of $AdS_3/CFT_2$ framework and later generalized to the $AdS_{d+1}/CFT_d$ scenario in \cite{Fursaev:2006ih,Casini:2011kv,Faulkner:2013yia,Lewkowycz:2013nqa}. Subsequently the covariant HRT conjecture was proved in \cite{Dong:2016hjy}. The developments described above naturally led to the interesting issue of a corresponding holographic characterization for the universal part of the entanglement negativity of $CFT_d$s in the $AdS_{d+1}/CFT_d$ scenario. A holographic computation of the entanglement negativity for the pure vacuum state of a $CFT_d$ dual to a bulk pure $AdS_{d+1}$ geometry was given in \cite{Rangamani:2014ywa}. Despite this progress a clear holographic construction for the entanglement negativity of bipartite states in $CFT_d$s remained an outstanding issue.

In \cite{Chaturvedi:2016rcn,Chaturvedi:2016opa} two of the present authors (VM and GS) proposed a holographic entanglement negativity conjecture and its covariant generalization for bipartite states in the $AdS_3/CFT_2$ scenario. This was substantiated by a large central charge analysis of the entanglement negativity of the $CFT_{1+1}$ utilizing the monodromy technique in \cite{Malvimat:2017yaj}. This proposal was subsequently extended in \cite{Chaturvedi:2016rft} to higher dimensions in the context of the $AdS_{d+1}/CFT_d$. However a bulk proof of this conjecture along the lines of \cite{Faulkner:2013yia,Lewkowycz:2013nqa} remains an outstanding issue. Following \cite{Chaturvedi:2016rcn,Chaturvedi:2016opa} in \cite{Jain:2017aqk,Jain:2017uhe} a holographic entanglement negativity conjecture and its covariant extension was proposed for bipartite mixed state configurations of adjacent intervals in dual $CFT_{1+1}$s. Subsequently through the $AdS_{d+1}/CFT_d$ framework a higher dimensional generalization of the above construction was proposed in \cite{Jain:2017xsu}. This could be applied to investigate such mixed states in $CFT_d$s dual to the bulk pure $AdS_{d+1}$ geometry, $AdS_{d+1}$-Schwarzschild black hole and the $AdS_{d+1}$-Reissner Nordstrom black hole in \cite{Jain:2017xsu,Jain:2018bai}.

As mentioned earlier the entanglement negativity for the mixed state of two disjoint intervals which is in general non universal exhibits an interesting behavior in the large central charge limit where a universal contribution may be isolated. A holographic description from a bulk perspective for this intriguing behavior of the entanglement negativity is a fascinating open issue. In this article we address this interesting issue and propose a holographic entanglement negativity conjecture for such mixed state configuration of two disjoint intervals in the $AdS_3/CFT_2$ scenario. To this end we utilize the large central charge analysis presented in \cite{Kulaxizi:2014nma} to extract the universal part of the entanglement negativity for the mixed state in question both at zero and finite temperatures and also for a finite size system in  $CFT_{1+1}$. Interestingly we observe that the entanglement negativity for the mixed states in question are cut of{}f independent. Following this analysis it is possible to establish a holographic conjecture characterizing the universal part of the entanglement negativity of the mixed state in question. Our construction involves a specific algebraic sum of the lengths of bulk space like geodesics anchored on intervals appropriate to the configuration of the mixed state in question and reduces to an algebraic sum of the holographic mutual informations between particular combinations of the intervals.\footnote{Our analysis has been confirmed in recent articles \cite {Shapourian:2018lsz,Kudler-Flam:2018qjo}.} Application of our conjecture to the examples of such mixed state configurations in $CFT_{1+1}$ dual to bulk pure $AdS_3$ geometries and the Euclidean BTZ black hole substantiates our conjecture and constitute significant consistency checks. Interestingly in the limit of the intervals being adjacent we are able to exactly reproduce the universal features of results described in \cite{Jain:2017aqk,Calabrese:2012nk,Calabrese:2014yza} from our holographic construction for the disjoint case.

This article is organized as follows. In section \ref{sec_en} we briefly review the computation of entanglement negativity for bipartite mixed state configuration of two disjoint intervals in a $CFT_{1+1}$. In section \ref{sec_en_large_c} we describe the large central charge analysis for the entanglement negativity utilizing the monodromy technique. Subsequently in section \ref{sec_hen_dj_int} we advance a holographic entanglement negativity conjecture for the mixed state of disjoint intervals using the large central charge results and describe its application to various scenarios. Finally, we summarize our results in section \ref{sec_summary} and present our conclusions.

\section{Entanglement negativity}
\label{sec_en}

We begin with an outline of the salient features of entanglement negativity in quantum information theory \cite{Vidal:2002zz} (for a brief review also see \cite{Rangamani:2014ywa}). In this context it is necessary to consider a tripartite system in a pure state constituted by the subsystems $ A_1 $, $ A_2 $ and $ B $. The bipartite system $ A \equiv A_1 \cup A_2 $ in a mixed state, described by the reduced density matrix $ \rho_A = \operatorname{Tr}_B (\rho) $, may then be obtained by tracing over the subsystem $ B \equiv A^c $. It is assumed that the Hilbert space for the bipartite system $ A $ may be expressed as a direct product $ {\cal H} = {\cal H}_1 \otimes {\cal H}_2 $ where $ {\cal H}_1 $ and $ {\cal H}_2 $ respectively describe the Hilbert spaces for the subsystems $ A_1 $ and $ A_2 $. The partial transpose of the reduced density matrix $ \rho_A $ with respect to $ A_2 $, is defined as
\begin{equation}
\label{eq_def_prtl_transps}
\left \langle e^{(1)}_i e^{(2)}_j \middle | \rho_A^{T_2}
\middle | e^{(1)}_k e^{(2)}_l \right \rangle
= \left \langle e^{(1)}_i e^{(2)}_l \middle | \rho_A
\middle | e^{(1)}_k e^{(2)}_j \right \rangle ,
\end{equation}
where $ \lvert e^{(1)}_i \rangle $ and $ \lvert e^{(2)}_j \rangle $ represent the bases for $ {\cal H}_1 $ and $ {\cal H}_2 $ respectively. The entanglement negativity $ {\cal E} $ which characterizes the entanglement between the subsystems $ A_1 $ and $ A_2 $ may then be defined as follows
\begin{equation}
\label{eq_def_ent_neg}
{\cal E}
= \ln \left \lVert \rho_A^{T_2} \right \rVert,
\end{equation}
where $\left \lVert \rho_A^{T_2} \right \rVert$ is the trace norm of the matrix $\rho_A^{T_2}$.

\subsection{Entanglement negativity in a $CFT_{1+1}$}
\label{subsec_en_cft_2}

\begin{figure}[H]
\begin{center}
\includegraphics[scale=1.5]{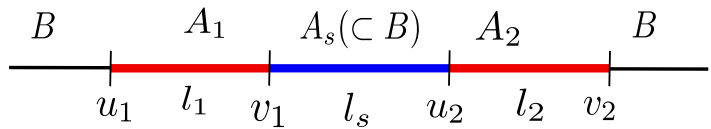}
\caption{Tripartite system of two disjoint intervals $ A_1 $ and $ A_2 $, and the remainder of the system $B$ in a $CFT_{1+1}$. The two intervals are separated by $ A_s $.}
\label{fig_en_dj_int}
\end{center}
\end{figure}

A systematic procedure to compute the entanglement negativity for bipartite states in a $CFT_{1+1}$, utilizing a replica technique, was described in \cite{Calabrese:2012ew,Calabrese:2012nk,Calabrese:2014yza}. The entanglement negativity in this case involves the quantity $ \operatorname{Tr} \left ( \rho_A^{T_2} \right )^{n_e} $ for even $ n_e $ and the analytic continuation of even sequences of $ n_e $ to $ n_e \to 1 $ leads to the following expression
\begin{equation}
\label{eq_ent_neg_replica_limit}
{\cal E} = \lim_{ n_e \to 1 } \ln
\operatorname{Tr} \left ( \rho_A^{T_2} \right )^{n_e} .
\end{equation}
The specific mixed state configuration under consideration is described by the disjoint intervals $ A_1 \equiv \left [ u_1, v_1 \right ] $ (of length $ l_1 $) and $ A_2 \equiv \left [ u_2, v_2 \right ] $ (of length $ l_2 $) with $ A \equiv A_1 \cup A_2 $ while $ B \equiv A^c $ denotes the rest of the system as shown in figure \ref{fig_en_dj_int}. The interval $ \left [ v_1, u_2 \right ] $ (of length $ l_s $) separating $ A_1 $ and $ A_2 $ in this case is described by $ A_s \subset B $.
%
% ===== ===== ===== Figure ===== ===== ===== 
%
%
% ===== ===== ===== ===== ===== ===== ===== 
%
For this configuration, the quantity $ \operatorname{Tr} \left ( \rho_A^{T_2} \right )^{n_e} $ may be expressed in terms of a four point twist correlator on the complex plane $ \mathbb{C} $ as
\begin{equation}
\label{eq_4pt_fn}
\operatorname{Tr} \left ( \rho_A^{T_2} \right )^{n_e}
= \left \langle {\cal T}_{n_e} (u_1)
\overline{{\cal T}}_{n_e} (v_1)
\overline{{\cal T}}_{n_e} (u_2)
{\cal T}_{n_e} (v_2) \right \rangle_{\mathbb{C}} .
\end{equation}
As described in \cite{Calabrese:2012nk} the above four point twist correlator may be expressed in the replica limit $ n_e \to 1 $ as follows
\begin{equation}
\label{eq_4pt_fn_expr_replica_lim}
\lim_{n_e \to 1}
\left \langle {\cal T}_{n_e} (u_1)
\overline{{\cal T}}_{n_e} (v_1)
\overline{{\cal T}}_{n_e} (u_2)
{\cal T}_{n_e} (v_2) \right \rangle_{\mathbb{C}}
= {\cal G}(x),
\end{equation}
where $ {\cal G}(x) $  is a non universal function of the cross ratio
$ x = \left [ \left ( v_1 - u_1 \right ) \left ( v_2 - u_2 \right ) \right ] /
\left [ \left ( u_2 - u_1 \right ) \left ( v_2 - v_1 \right ) \right ] $
and depends on the full operator content of the corresponding $CFT_{1+1}$. In the next section, we will review the computation of an explicit universal form of this function in the large central charge limit.

\section{Entanglement negativity at large $ c $}
\label{sec_en_large_c}

In this section we  briefly review the large central charge analysis for the four point twist correlator in eq.\@ \eqref{eq_4pt_fn} above through the monodromy technique \cite{Belavin:1984vu,1987TMP....73.1088Z,Zamolodchikov:1995aa,Harlow:2011ny,Hartman:2013mia,Fitzpatrick:2014vua,Kulaxizi:2014nma,Alkalaev:2015lca,Fitzpatrick:2015zha,Perlmutter:2015iya,Ruggiero:2018hyl}. Our discussion will be focused on the explicit form of the four point twist correlator when the disjoint intervals depicted in figure \ref{fig_en_dj_int} are in proximity as described in \cite{Kulaxizi:2014nma}. In this instance the entanglement negativity for the bipartite zero temperature mixed state configuration of disjoint intervals may be obtained explicitly in the large central charge limit.

For this purpose it is required to analyze a four point correlation function of primary fields in a $CFT_{1+1}$ which is given as
$ \left \langle {\cal O}_1 (z_1)
{\cal O}_2 (z_2) {\cal O}_3 (z_3)
{\cal O}_4 (z_4) \right \rangle_{\mathbb{C}} $.
It is possible to set $ z_1 = 0, z_2 = x, z_3 = 1, z_4 = \infty $ through the conformal transformation
$ z \to \left [ (z_1 - z) (z_3 - z_4) \right ] /$
$\left [ (z_1 - z_3) (z - z_4) \right ] $,
where $ x \equiv ( z_{12} z_{34} ) / ( z_{13} z_{24} ) $
is the relevant cross ratio with $ z_{ij} \equiv z_i - z_j $. The resulting four point correlator may then be expanded in terms of the conformal blocks as follows
\begin{equation}
\label{eq_4pt_fn_conf_block_expans}
\begin{aligned}
& \left \langle {\cal O}_1 (0)
{\cal O}_2 (x) {\cal O}_3 (1)
{\cal O}_4 (\infty) \right \rangle_{\mathbb{C}} \\
& = \sum_p a_p {\cal F} \left ( c, h_p, h_i, x \right )
\overline{{\cal F}} \left ( c, \bar h_p, \bar h_i, \bar x \right ) ,
\end{aligned}
\end{equation}
%
%\begin{equation}
%\label{eq_4pt_fn_conf_block_expans}
%\left \langle {\cal O}_1 (0)
%{\cal O}_2 (x) {\cal O}_3 (1)
%{\cal O}_4 (\infty) \right \rangle_{\mathbb{C}}
%= \sum_p a_p {\cal F} \left ( c, h_p, h_i, x \right )
%\overline{{\cal F}} \left ( c, \bar h_p, \bar h_i, \bar x \right ) ,
%\end{equation}
%
where we sum over all the primary operators $ {\cal O}_p $ with conformal dimensions $ \left ( h_p, \bar h_p \right ) $, and $ \left ( h_i, \bar h_i \right ) $ are the conformal dimensions of the operators $ {\cal O}_i $. An analytic expression for $ {\cal F} \left (c, h_p, h_i, x \right ) $ is not known except for some specific values of the parameters. However, in the semi classical approximation given by the large central charge limit $ c \to \infty $ with $ h_p / c $ and $ h_i / c $ fixed, the conformal block exponentiates \cite{Belavin:1984vu,1987TMP....73.1088Z} as follows
\begin{equation}
\label{eq_conf_block_large_c_approx}
{\cal F} \left (c, h_p, h_i, x \right )
\simeq \exp \left [
- \frac{ c }{ 6 }
f \left ( h_p / c, h_i / c, x \right )
\right ] .
\end{equation}
The function $ f $ in the above expression may then be determined through the monodromy properties of the solutions to the second order dif{}ferential equation given as
\begin{equation}
\label{eq_mono_tech_diff_eq}
\psi''(z) + T(z) \psi(z) = 0 ,
\end{equation}
around the above specified points $(0,x,1,\infty)$. In the above equation
\begin{equation}
\label{eq_stress_ten_expr}
T(z) = \sum_{i = 1}^4
\left [
\frac{ 6 h_i }{ c \left ( z - z_i \right )^2 }
- \frac{ c_i }{ z - z_i }
\right ] ,
\end{equation}
where $ c_i \equiv \partial f / \partial z_i $ are known as the accessory parameters. Three of these parameters, $ c_1 $, $ c_3 $ and $ c_4 $, may be fixed by the asymptotic form of $ T(z) \sim z^{ - 4 } $ as $ z \to \infty $, as follows
\begin{equation}
\label{eq_access_paramtr_cond}
\begin{gathered} 
\begin{aligned}
\sum \limits_{i=1}^4 c_i & = 0 ,
& \sum \limits_{i=1}^4 \left ( c_i z_i - \frac{ 6 h_i }{ c } \right ) & = 0,
\end{aligned} \\ 
\sum \limits_{i=1}^4 \left ( c_i z_i^2 
- \frac{ 12 h_i }{ c } z_i \right ) = 0 .
\end{gathered}
\end{equation}
%
%\begin{align}
%\label{eq_access_paramtr_cond}
%\sum \limits_{i=1}^4 c_i & = 0 ,
%& \sum \limits_{i=1}^4 \left ( c_i z_i - \frac{ 6 h_i }{ c } \right ) & = 0 ,
%& \sum \limits_{i=1}^4 \left ( c_i z_i^2 
%- \frac{ 12 h_i }{ c } z_i \right ) & = 0 .
%\end{align}
%
The remaining parameter $ c_2 $ is then determined by the monodromy condition which requires the trace of the monodromy matrix around a closed path enclosing the singularities of $ T(z) $, to satisfy
\begin{align}
\label{eq_mono_cond}
\operatorname{Tr} (M) & = - 2 \cos \left ( \pi \Lambda_p \right ) ,
& \Lambda_p & = \sqrt{ 1 - \frac{ 24 h_p }{ c } } ,
\end{align}
where $ h_p $ describes the lowest conformal dimension for the intermediate operator in the channel under consideration. The function $ f $ in eq.\@ \eqref{eq_conf_block_large_c_approx} may then be determined through the integration of the expression $ \partial f / \partial x = c_2 $.

\subsection{Entanglement negativity of disjoint intervals in the $ x \to 1 $ {channel}}
\label{subsec_en_dj_int}

We now describe the application of the monodromy technique to the four point twist correlator characterizing the entanglement negativity for the $ x \to 1 $ channel in which the intervals are in close proximity to each other \cite{Kulaxizi:2014nma}. To go to the complex plane, we make the following identification: $ \left ( u_1, v_1, u_2, v_2 \right ) \equiv \left ( z_1, z_2, z_3, z_4 \right ) $. As mentioned earlier the four point twist correlator in eq.\@ \eqref{eq_4pt_fn} admits the following conformal block expansion given as
\begin{equation}
\label{eq_twist_4pt_fn_conf_block_expans}
\begin{aligned}
& \left \langle {\cal T}_{n_e} (z_1)
\overline{{\cal T}}_{n_e} (z_2)
\overline{{\cal T}}_{n_e} (z_3)
{\cal T}_{n_e} (z_4) \right \rangle_{\mathbb{C}} \\
& = \sum_p a_p {\cal F} \left ( c, h_p, h_i, x \right )
\overline{{\cal F}} \left ( c, \bar h_p, \bar h_i, \bar x \right ) .
\end{aligned}
\end{equation}
%
%\begin{equation}
%\label{eq_twist_4pt_fn_conf_block_expans}
%\left \langle {\cal T}_{n_e} (z_1)
%\overline{{\cal T}}_{n_e} (z_2)
%\overline{{\cal T}}_{n_e} (z_3)
%{\cal T}_{n_e} (z_4) \right \rangle_{\mathbb{C}}
%= \sum_p a_p {\cal F} \left ( c, h_p, h_i, x \right )
%\overline{{\cal F}} \left ( c, \bar h_p, \bar h_i, \bar x \right ) .
%\end{equation}
%

In the limit $ x \to 1 $ where the disjoint intervals are in close proximity, the relevant intermediate operator has been shown to be $ \overline{{\cal T}}_{n_e}^2 $ \cite{Kulaxizi:2014nma}. Hence the dominant contribution to the four point twist correlator in eq.\@ \eqref{eq_twist_4pt_fn_conf_block_expans} arises from the conformal block with the conformal dimension $ h_p = h_{\overline{{\cal T}}_{n_e}^2} \equiv \hat h $. In the large $ c $ limit, using eq.\@ \eqref{eq_conf_block_large_c_approx} we then arrive at the following
\begin{equation}
\label{eq_twist_4pt_fn_conf_block_large_c_approx}
\begin{aligned}
& \left \langle {\cal T}_{n_e} (z_1)
\overline{{\cal T}}_{n_e} (z_2)
\overline{{\cal T}}_{n_e} (z_3)
{\cal T}_{n_e} (z_4) \right \rangle_{\mathbb{C}} \\
& \simeq {\cal F} \left ( c, h_p = \hat{h}, h_i, x \right )
\overline{{\cal F}} \left ( c, \bar h_p = \hat{h}, \bar h_i, \bar x \right ) \\
& \simeq \exp \left [ - \frac{ c }{ 3 }
f \left ( \hat h / c, h_i / c, x \right ) \right ] .
\end{aligned}
\end{equation}
%
%\begin{equation}
%\label{eq_twist_4pt_fn_conf_block_large_c_approx}
%\begin{aligned}
%\left \langle {\cal T}_{n_e} (z_1)
%\overline{{\cal T}}_{n_e} (z_2)
%\overline{{\cal T}}_{n_e} (z_3)
%{\cal T}_{n_e} (z_4) \right \rangle_{\mathbb{C}}
%& \simeq {\cal F} \left ( c, h_p = \hat{h}, h_i, x \right )
%\overline{{\cal F}} \left ( c, \bar h_p = \hat{h}, \bar h_i, \bar x \right ) \\
%& \simeq \exp \left [ - \frac{ c }{ 3 }
%f \left ( \hat h / c, h_i / c, x \right ) \right ] .
%\end{aligned}
%\end{equation}
%
In the replica limit $ n_e \to 1 $ we have $ h_i = 0 $ and $ \hat h = - c / 8 $.\footnote{Note that the negative value of scaling dimension of the twist field in the replica limit has to be understood as an analytic continuation.} In this case $ T(z) $ in eq.\@ \eqref{eq_stress_ten_expr} is given by
\begin{equation}
\label{eq_stress_ten_replica_large_c}
T(z) \simeq \frac{ c_2 (1 - x) }{ (z - 1)^2 } ,
\end{equation}
hence the dif{}ferential equation in eq.\@ \eqref{eq_mono_tech_diff_eq} reduces to
\begin{equation}
\label{eq_mono_tech_diff_eq_replica_large_c}
\psi''(z) + \frac{ c_2 (1 - x) }{ (z - 1)^2 } \psi(z) = 0 .
\end{equation}
Requiring the solutions of the above dif{}ferential equation in eq.\@ \eqref{eq_mono_tech_diff_eq_replica_large_c} to satisfy the monodromy condition mentioned in  eq.\@ \eqref{eq_mono_cond}, the accessory parameter $ c_2 $ may be determined as
\begin{equation}
\label{eq_c2_expr}
c_2 = - \frac{ 3 }{ 4 } \left ( \frac{ 1 }{1 - x} \right ) .
\end{equation}
Using the above expression for the accessory parameter $ c_2 $ we may now obtain the function $ f $ in eq.\@ \eqref{eq_twist_4pt_fn_conf_block_large_c_approx} as follows\footnote{Note that in computing $ f $ from eq.\@ \eqref{eq_c2_expr}, the integration constant has been set to zero in accordance with the result obtained in \cite{Calabrese:2012nk} in the limit  $ x \to 1 $.}
\begin{equation}
\label{eq_f_expr}
f = \frac{ 3 }{ 4 } \ln \left ( 1 - x \right ) .
\end{equation}
The four point twist correlator in eq.\@ \eqref{eq_twist_4pt_fn_conf_block_large_c_approx} may now be given in the large $c$ limit as
\begin{equation}
\label{eq_twist_4pt_fn_ito_x}
\lim_{ n_e \to 1 }
\left \langle {\cal T}_{n_e} (z_1)
\overline{{\cal T}}_{n_e} (z_2)
\overline{{\cal T}}_{n_e} (z_3)
{\cal T}_{n_e} (z_4) \right \rangle_{\mathbb{C}}
= \left ( 1 - x \right )^{ 2 \hat h } .
\end{equation}
Utilizing eqs.\@ \eqref{eq_ent_neg_replica_limit} and \eqref{eq_4pt_fn}, the entanglement negativity for the bipartite mixed state configuration of disjoint intervals in proximity may now be obtained from eq.\@ \eqref{eq_twist_4pt_fn_ito_x} upon substitution of the cross ratio
$ x \equiv ( z_{12} z_{34} ) / ( z_{13} z_{24} ) $ as follows
\begin{equation}
\label{eq_ent_neg_expr_dj_int}
{\cal E} = \frac{ c }{ 4 }
\ln \left (
\frac{ |z_{13}| |z_{24}| }
{ |z_{14}| |z_{23}| }
\right ) .
\end{equation}
The above equation provides a general expression for the entanglement negativity  of the bipartite mixed state configuration of disjoint intervals when they are in proximity to each other in the large central charge limit. We will now proceed to utilize the above expression to obtain the entanglement negativity of the mixed state in question for dif{}ferent scenarios and demonstrate that they match with the established results in the adjacent limit \cite{Calabrese:2012nk} at large central charge.

Note that in \cite{Hartman:2013mia}, it was demonstrated  that in the large central charge limit, the entanglement entropy of the configuration described by two disjoint intervals exhibits a phase transition from its value in the $ s $-channel ($ x \to 0 $) to its value in the $ t $-channel ($ x \to 1 $) at $ x = \frac{ 1 }{ 2 } $ . These correspond to dif{}ferent geodesic combinations in the dual bulk $AdS_3$ geometry as predicted by the holographic proposal of Ryu and Takayanagi. Interestingly, in \cite{Kulaxizi:2014nma}, the authors showed that a similar phase transition occurs for the entanglement negativity of the mixed state of two disjoint intervals as well. It was numerically  demonstrated  that this phase transition occurs for the  negativity in the large $ c $ limit  from its value in the $ s $-channel ($ x \to 0 $) where it vanishes, to its value in the $ t $-channel which is given by eq.\@ \eqref{eq_ent_neg_expr_dj_int}. However, it was not possible to determine the exact value of the cross ratio $ x $ at which the phase transition occurs. Recently, in \cite{Dong:2018esp} it was shown that there exists a correspondence between the classical geometries  dual to the R\'{e}nyi entanglement entropy and the R\'{e}nyi entanglement negativity which suggests that the phase transition once again occurs at $ x = \frac{ 1 }{ 2 } $.

\subsection{Entanglement negativity for disjoint intervals in vacuum at large $ c $}
\label{subsec_en_dj_int_vac}

The general expression described above in eq.\@ \eqref{eq_ent_neg_expr_dj_int} may now be employed to obtain the entanglement negativity for the zero temperature mixed state of two disjoint intervals in proximity through the substitution of the lengths of the respective intervals, leading to the following expression  
\begin{equation}
\label{eq_ent_neg_dj_int_vac}
{\cal E} = \frac{ c }{ 4 }
\ln \left [ \frac{ \left ( l_1 + l_s \right ) \left ( l_2 + l_s \right ) }
{ l_s \left ( l_1 + l_2 + l_s \right ) } \right ] .
\end{equation}
Note that the above expression describes the universal part of the entanglement negativity for the zero temperature mixed state under consideration in the large central charge limit.

Interestingly the above result is cut of{}f independent unlike the case for the mixed state of adjacent intervals as described in \cite{Calabrese:2012nk}. Furthermore it is to be noted that the above expression in eq.\@ \eqref{eq_ent_neg_dj_int_vac} exactly reproduces the universal part of the entanglement negativity in the adjacent interval limit provided the separation length $ l_s $ is identified with the cut of{}f as $ l_s \to a $  \cite{Calabrese:2012nk}.

\subsection{Entanglement negativity for disjoint intervals in vacuum for a finite size system at large $ c $}
\label{subsec_en_dj_int_vac_fin_size}

For a finite size system of length $ L $ with a periodic boundary condition, the entanglement negativity for the mixed state in question, may be obtained from eq.\@ \eqref{eq_ent_neg_expr_dj_int} through the conformal transformation $ z \to w = \left ( i L / 2 \pi \right ) \ln z $, from the complex plane to the cylinder of circumference $ L $ \cite{Calabrese:2012nk}. Under this conformal map the four point twist correlator in eq.\@ \eqref{eq_4pt_fn} transforms as
\begin{equation}
\label{eq_twist_4pt_fn_trfn}
\begin{aligned}
& \left \langle {\cal T}_{n_e} (w_1)
\overline{{\cal T}}_{n_e} (w_2)
\overline{{\cal T}}_{n_e} (w_3)
{\cal T}_{n_e} (w_4) \right \rangle_{cyl} \\
& = \prod_{i=1}^{4}
\left [ \left ( \frac{ d w(z) }{ d z } \right )^{ - \Delta_i }
\right ]_{ z = z_i } \\
& \times \left \langle {\cal T}_{n_e} (z_1)
\overline{{\cal T}}_{n_e} (z_2)
\overline{{\cal T}}_{n_e} (z_3)
{\cal T}_{n_e} (z_4) \right \rangle_{ \mathbb{C} } ,
\end{aligned}
\end{equation}
%
%\begin{equation}
%\label{eq_twist_4pt_fn_trfn}
%\left \langle {\cal T}_{n_e} (w_1)
%\overline{{\cal T}}_{n_e} (w_2)
%\overline{{\cal T}}_{n_e} (w_3)
%{\cal T}_{n_e} (w_4) \right \rangle_{cyl}
%= \prod_{i=1}^{4}
%\left [ \left ( \frac{ d w(z) }{ d z } \right )^{ - \Delta_i }
%\right ]_{ z = z_i }
%\left \langle {\cal T}_{n_e} (z_1)
%\overline{{\cal T}}_{n_e} (z_2)
%\overline{{\cal T}}_{n_e} (z_3)
%{\cal T}_{n_e} (z_4) \right \rangle_{ \mathbb{C} } ,
%\end{equation}
%
where $ \Delta_i $ are the scaling dimensions of the twist fields at the locations $ w = w_i $.

Utilizing eqs.\@ \eqref{eq_ent_neg_replica_limit}, \eqref{eq_4pt_fn} and \eqref{eq_twist_4pt_fn_trfn}, the entanglement negativity at large $ c $ for the zero temperature mixed state configuration of disjoint intervals in proximity for this case is then obtained from eq.\@ \eqref{eq_twist_4pt_fn_ito_x} as follows
\begin{equation}
\label{eq_ent_neg_dj_int_finite_size}
{\cal E} = \frac{ c }{ 4 }
\ln \left [
\frac
{
\sin \frac{ \pi \left ( l_1 + l_s \right ) }{ L }
\sin \frac{ \pi \left ( l_2 + l_s \right ) }{ L }
}{
\sin \frac{ \pi l_s }{ L }
\sin \frac{ \pi \left ( l_1 + l_2 + l_s \right ) }{ L }
}
\right ] .
\end{equation}
Note that this result is also cut of{}f independent, in contrast to the case of adjacent intervals \cite{Calabrese:2012nk}. Once more the above expression exactly reproduces the corresponding entanglement negativity at large central charge for adjacent intervals in the limit $ l_s \to a $.

\subsection{Entanglement negativity for disjoint intervals at a finite temperature at large $ c $}
\label{subsec_en_dj_int_fin_temp}

For the mixed state in question at a finite temperature $ T $, the entanglement negativity at large $c$ may be obtained as above through the conformal map $ z \to w = \left ( \beta / 2 \pi \right ) \ln z $ from the complex plane to the cylinder where the Euclidean time direction has now been compactified to a circle with circumference $ \beta \equiv 1 / T $ \cite{Calabrese:2014yza}. As before, employing eqs.\@ \eqref{eq_ent_neg_replica_limit}, \eqref{eq_4pt_fn} and \eqref{eq_twist_4pt_fn_trfn}, with the transformation described above, the entanglement negativity at large $ c $, for the mixed state configuration of disjoint intervals in proximity at a finite temperature may be computed from eq.\@ \eqref{eq_twist_4pt_fn_ito_x} as follows
\begin{equation}
\label{eq_ent_neg_dj_int_finite_temp}
{\cal E} = \frac{ c }{ 4 }
\ln \left [
\frac
{
\sinh \frac{ \pi \left ( l_1 + l_s \right ) }{ \beta }
\sinh \frac{\pi \left ( l_2 + l_s \right ) }{ \beta }
}{
\sinh \frac{ \pi l_s }{ \beta }
\sinh \frac{ \pi \left ( l_1 + l_2 + l_s \right ) }{ \beta }
}
\right ] .
\end{equation}
As earlier this result is also cut of{}f independent and reproduces the corresponding large central charge result for adjacent intervals \cite{Calabrese:2014yza,Jain:2017aqk} in the limit $ l_s \to a $.

\begin{figure}[H]
\centering
\begin{subfigure}{0.5\textwidth}
  \centering
  \includegraphics[width=0.95\linewidth]{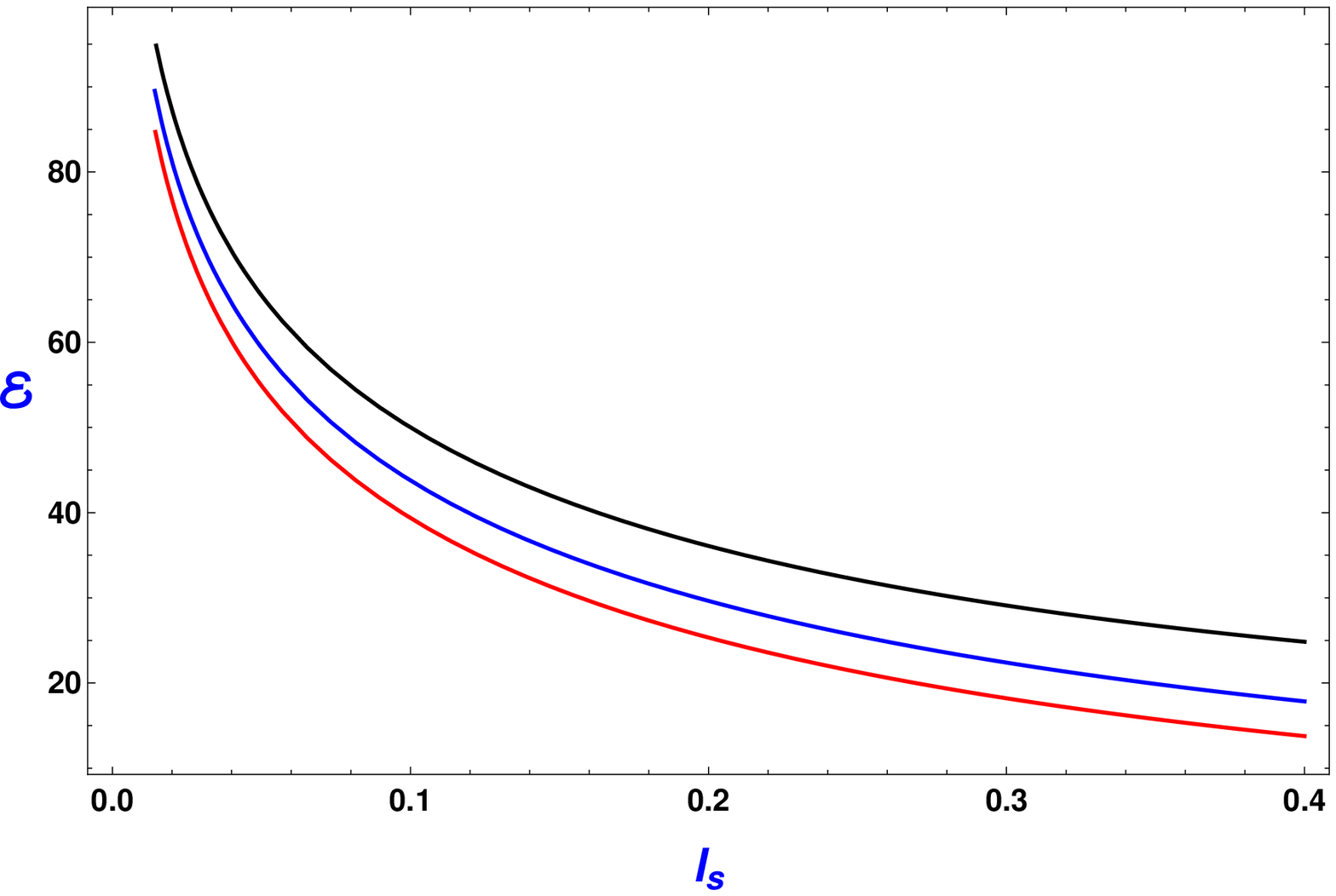}
  \caption{$ {\cal E} $ vs.\@ $ l_s $ plots}
  \label{fig_plot_en_vs_ls}
\end{subfigure}%
\begin{subfigure}{.5\textwidth}
  \centering
  \includegraphics[width=0.95\linewidth]{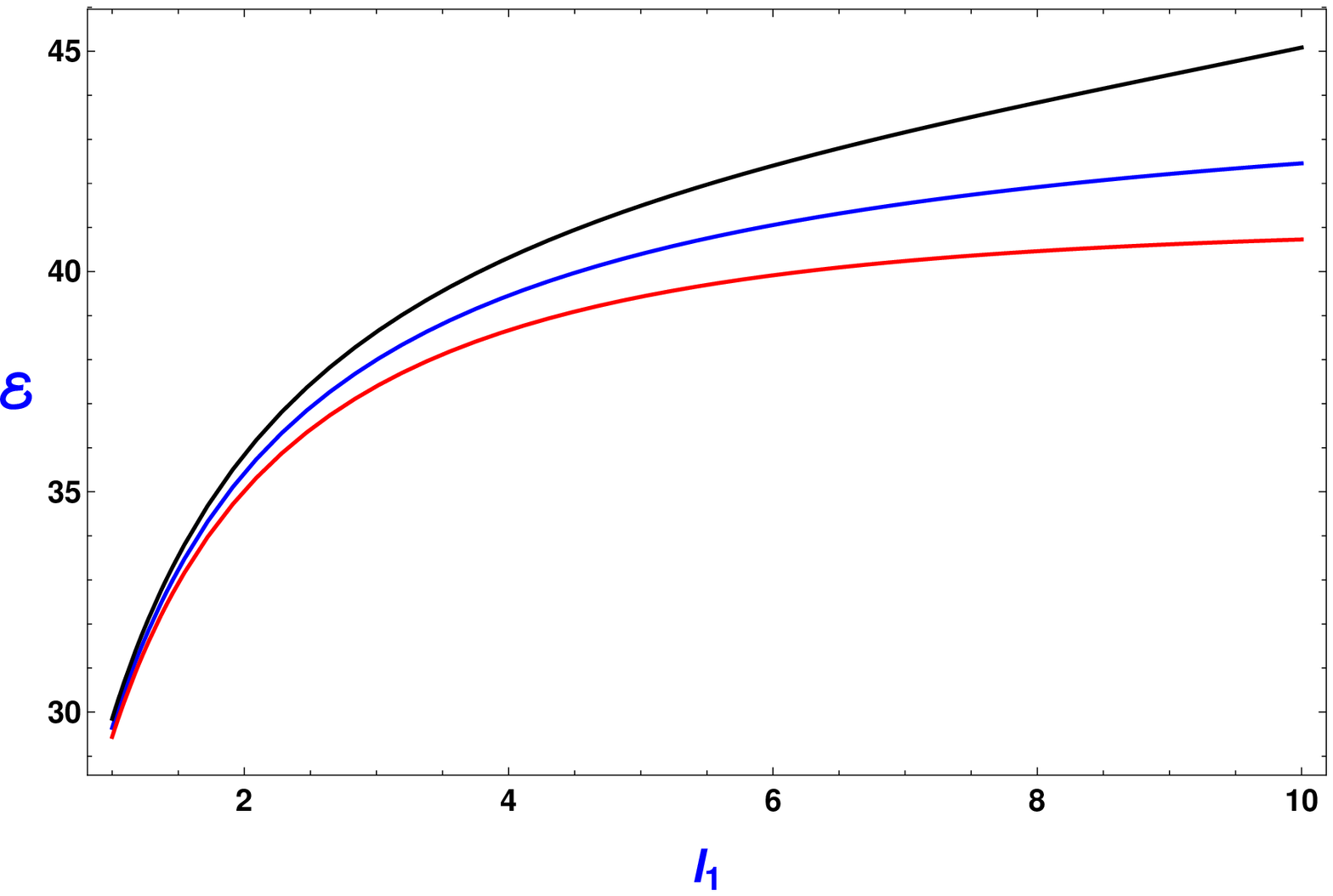}
  \caption{$ {\cal E} $ vs.\@ $ l_1 $ plots}
  \label{fig_plot_en_vs_l1}
\end{subfigure}
\caption{Negativity plots for two disjoint intervals in a $CFT_{1+1}$ in dif{}ferent cases. In both the figures, the blue, black and red curves represent the vacuum scenario, finite size system and finite temperature system, respectively. Figure (a) shows plots of negativity vs.\@ separation between the intervals. Here, $ c = 100, l_1 = l_2 = 1, L = \beta = 4 $.
Figure (b) presents plots of negativity vs.\@ length of the first interval. In this case, $c = 100, l_2 = 1, l_s = 0.2, L = \beta = 20 $.}
\label{fig_plot_ent_neg}
\end{figure}

In figure \ref{fig_plot_en_vs_ls} we graphically describe the behavior of the entanglement negativity as a function of the separation $ l_s $ between the disjoint intervals for the three scenarios described above. It is observed in all the cases that the entanglement negativity decreases as we increase separation length $ l_s $ between the intervals, which is in conformity with quantum information results. In figure \ref{fig_plot_en_vs_l1} the entanglement negativity has been plotted against the length of the first interval $l_1$. In this plot we observe that the entanglement negativity increases with the interval size and eventually saturates for large $ l_1 $ in all the cases.

%
% ===== ===== ===== Figure ===== ===== ===== 
%
%
% ===== ===== ===== ===== ===== ===== ===== 
%

Having presented the entanglement negativity of the mixed state under consideration in a $CFT_{1+1}$ for the three dif{}ferent scenarios we now proceed to establish a holographic conjecture involving the dual bulk $AdS_3$ geometry which correctly reproduces the above large central charge results.

\section{Holographic entanglement negativity for disjoint intervals}
\label{sec_hen_dj_int}

In this section we establish a holographic entanglement negativity conjecture in the $AdS_3/CFT_2$ framework for the mixed states of disjoint intervals in proximity from the large central charge results obtained in the previous section. As earlier we consider the disjoint intervals $ A_1 $ and $ A_2 $ of lengths $ l_1 $ and $ l_2 $ respectively depicted in figure \ref{fig_en_dj_int}, where the subsystem $ A \equiv A_1 \cup A_2 $ is in a mixed state. The separation between the intervals corresponds to the subsystem $ A_s \subset B $ of length $ l_s $ where $B=A^c$ denotes the rest of the system. We now consider the two point twist correlator in a holographic $CFT_{1+1}$ which is given as
\begin{equation}
\label{eq_twist_2pt_fn_cft}
\left \langle {\cal T}_{n_e} (z_i)
\overline{{\cal T}}_{n_e} (z_j) \right \rangle_{\mathbb{C}}
\sim \left | z_{ij} \right |^{- 2 \Delta_{ {\cal T}_{n_e}} } .
\end{equation}
According to the $AdS_3/CFT_2$ dictionary, the above two point twist correlator may be expressed in terms of the length $ {\cal L}_{ij} $ of the bulk space like geodesic anchored on the corresponding interval (in the geodesic approximation) \cite{Ryu:2006ef} as follows
\begin{equation}
\label{eq_twist_2pt_fn_ads_cft}
\left \langle {\cal T}_{n_e} (z_i)
\overline{{\cal T}}_{n_e} (z_j) \right \rangle_{\mathbb{C}}
\sim
\exp \left ( - \frac{ \Delta_{{\cal T}_{n_e}} {\cal L}_{ij} }{ R } \right ) ,
\end{equation}
where $ R $ is the $AdS_3$ length scale. 

Utilizing eqs.\@ \eqref{eq_twist_2pt_fn_cft} and \eqref{eq_twist_2pt_fn_ads_cft}, the four point twist correlator in the holographic $CFT_{1+1}$ as given in eq.\@ \eqref{eq_twist_4pt_fn_ito_x} may be expressed in terms of the lengths of the bulk space like geodesics as follows
\begin{equation}
\label{eq_twist_4pt_fn_ito_geo_len}
\begin{aligned}
& \lim_{ n_e \to 1 }
\left \langle {\cal T}_{n_e} (z_1)
\overline{{\cal T}}_{n_e} (z_2)
\overline{{\cal T}}_{n_e} (z_3)
{\cal T}_{n_e} (z_4) \right \rangle_{\mathbb{C}} \\
& = \exp \left [ \frac{ c }{ 8 R }
\left ( {\cal L}_{13} + {\cal L}_{24}
- {\cal L}_{14} - {\cal L}_{23} \right ) \right ] .
\end{aligned}
\end{equation}
%
%\begin{equation}
%\label{eq_twist_4pt_fn_ito_geo_len}
%\lim_{ n_e \to 1 }
%\left \langle {\cal T}_{n_e} (z_1)
%\overline{{\cal T}}_{n_e} (z_2)
%\overline{{\cal T}}_{n_e} (z_3)
%{\cal T}_{n_e} (z_4) \right \rangle_{\mathbb{C}}
%= \exp \left [ \frac{ c }{ 8 R }
%\left ( {\cal L}_{13} + {\cal L}_{24}
%- {\cal L}_{14} - {\cal L}_{23} \right ) \right ] .
%\end{equation}
%

Using eqs.\@ \eqref{eq_ent_neg_replica_limit}, \eqref{eq_4pt_fn} and \eqref{eq_twist_4pt_fn_ito_geo_len} we arrive at the following holographic description for the entanglement negativity of the mixed state in terms of the lengths of the bulk space like geodesics 
\begin{equation}
\label{eq_holo_ent_neg_dj_int_geo_len}
{\cal E} = \frac{ 3 }{ 16 G_N^{(3)} }
\left ( {\cal L}_{13} + {\cal L}_{24}
- {\cal L}_{14} - {\cal L}_{23} \right ) ,
\end{equation}
where we have used the Brown-Henneaux formula $ c = \frac{ 3 R }{ 2 G_N^{(3)} } $
\cite{Brown:1986nw}.

\begin{figure}[H]
\begin{center}
\includegraphics[scale=2]{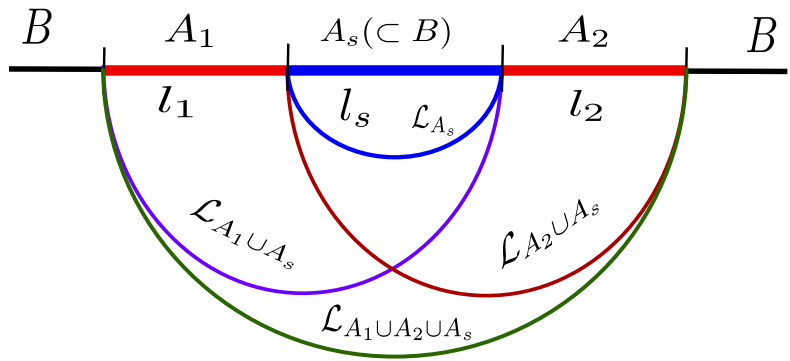}
\caption{Geodesics in a bulk $AdS_3$ anchored on dif{}ferent subsystems in the dual boundary $CFT_{1+1}$.}
\label{fig_hen_dj_int}
\end{center}
\end{figure}

It is observed from the above expression given by eq.\@ \eqref{eq_holo_ent_neg_dj_int_geo_len},  that the holographic entanglement negativity for the mixed state of disjoint intervals in proximity involves a specific algebraic sum of the lengths of the bulk space like geodesics anchored on the corresponding intervals as depicted in figure \ref{fig_hen_dj_int}. Interestingly, in the limit of adjacent intervals $l_s \to a$ ($ {\cal L}_{23} \to 0 $ in the bulk ), the above expression exactly reduces to the holographic entanglement negativity for the corresponding mixed state configuration described in \cite{Jain:2017aqk}. 
%
% ===== ===== ===== Figure ===== ===== ===== 
%
%
% ===== ===== ===== ===== ===== ===== ===== 
%
Note that as explained in subsection \ref{subsec_en_dj_int}, the result given in eq.\@ \eqref{eq_ent_neg_expr_dj_int} is valid for the values of the cross ratio $ \frac{ 1 }{ 2 } < x < 1 $ which implies that the geodesic combination given above is also valid only in this regime. On the other hand in the regime described by $ 0 < x < \frac{ 1 }{ 2 } $, the entanglement negativity is zero characterizing the phase transition at $x=\frac{1}{2}$. 

Utilizing the Ryu-Takayanagi conjecture \cite{Ryu:2006bv} for the holographic entanglement entropy of a subsystem $ \gamma $, given as $ S_{\gamma} = \frac{ \cal{L}_{\gamma} }{ 4 G_N^{(3)} } $, where $ \cal{L}_{\gamma} $ is the length of the space like geodesic anchored on the subsystem, we may express eq.\@ \eqref{eq_holo_ent_neg_dj_int_geo_len} as follows
\begin{equation}
\label{eq_holo_ent_neg_dj_int_holo_ent_enpy}
{\cal E} = \frac{ 3 }{ 4 }
\left ( S_{ A_1 \cup A_s } + S_{ A_s \cup A_2 }
- S_{ A_1 \cup A_2 \cup A_s } - S_{ A_s } \right ) .
\end{equation}
Interestingly the above expression in eq.\@ \eqref{eq_holo_ent_neg_dj_int_holo_ent_enpy} for the entanglement negativity may be expressed in terms of the holographic mutual informations between  appropriate subsystems as\footnote{There seems to be an intriguing connection between the holographic entanglement negativity and the holographic mutual information although they are distinct quantities in quantum information theory. For the adjacent intervals they are identical and this is also reported in the literature in \cite{Coser:2014gsa,Wen:2015qwa}.}
\begin{equation}
\label{eq_holo_ent_neg_dj_int_mut_info}
{\cal E} = \frac{ 3 }{ 4 }
\left [ {\cal I} \left ( A_1 \cup A_s, A_2 \right )
- {\cal I} \left ( A_s, A_2 \right ) \right ] .
\end{equation}
Note that here the holographic mutual information between subsystems $ A_i $ and $ A_j $ is denoted by $ {\cal I} \left ( A_i,A_j \right ) \equiv S_{ A_i } + S_{ A_j } - S_{ A_i \cup A_j } $. In the limit $ A_s \to \emptyset $ we recover the holographic entanglement negativity for the mixed state of adjacent intervals in terms of the holographic mutual information between the subsystems $ A_1 $ and $ A_2 $ as described in \cite{Jain:2017aqk}. This serves as a strong consistency check for our conjecture.

\subsection{Holographic entanglement negativity for two disjoint intervals in vacuum}
\label{subsec_hen_dj_int_vac}

Having established our holographic entanglement negativity conjecture for the mixed state configuration of disjoint intervals we now proceed to apply our conjecture to various scenarios described in the context of the $CFT_{1+1}$ in section \ref{sec_en_large_c} as consistency checks. To this end we begin by computing the holographic entanglement negativity for the zero temperature mixed state configuration of disjoint intervals in the $CFT_{1+1}$ vacuum which is dual to a bulk pure $AdS_3$ space time. This bulk geometry may be described in the Poincar\'{e} coordinates as follows
\begin{equation}
\label{eq_metric_pure_ads3_poincare_coord}
ds^2 = \left ( \frac{ r^2 }{ R^2 } \right )
\left ( - dt^2 + dx^2 \right )
+ \left ( \frac{ r^2 }{ R^2 } \right )^{ - 1 } dr^2 ,
\end{equation}
where $ R $ is the $AdS_3$ radius. The length of the bulk space like geodesic anchored on an interval $ \gamma $ ( of length $ l_{\gamma} $), in this geometry described by eq.\@ \eqref{eq_metric_pure_ads3_poincare_coord}, may then be expressed as \cite{Ryu:2006bv,Ryu:2006ef,Cadoni:2009tk,Cadoni:2010kla}
\begin{equation}
\label{eq_geo_len_pure_ads3_poincare_coord}
{\cal L}_{\gamma} = 2 R \ln \left ( \frac{ l_{\gamma} }{ a } \right ) ,
\end{equation}
with $ a $ being the UV cut of{}f. Using the expression in eq.\@ \eqref{eq_geo_len_pure_ads3_poincare_coord}, the holographic entanglement negativity for the mixed state under consideration may now be obtained from eq.\@ \eqref{eq_holo_ent_neg_dj_int_geo_len} as follows
\begin{equation}
\label{eq_holo_ent_neg_dj_int_vac}
{\cal E} = \frac{ 3 R }{ 8 G_N^{(3)} }
\ln \left [
\frac
{
\left ( l_1 + l_s \right ) \left ( l_2 + l_s \right )
}{
l_s \left ( l_1 + l_2 + l_s \right )
}
\right ] .
\end{equation}
Interestingly upon utilizing the Brown-Henneaux formula \cite{Brown:1986nw} our conjecture exactly reproduces the corresponding entanglement negativity for the disjoint intervals obtained through the replica technique in the large central charge limit given in eq.\@ \eqref{eq_ent_neg_dj_int_vac}. This serves as a consistency check for our conjecture. Furthermore in the adjacent limit $ l_s \to a $ where the separation between the intervals vanish, we recover the holographic entanglement negativity for the zero temperature mixed state of adjacent intervals described in \cite{Jain:2017aqk}. This also constitutes a validation of our construction.

\subsection{Holographic entanglement negativity of disjoint intervals in vacuum for a finite size system}
\label{subsec_hen_dj_int_vac_fin_size}

Having computed the vacuum entanglement negativity for the mixed state configuration of disjoint intervals, we now obtain the holographic entanglement negativity of the mixed state in question for a finite size system of length $ L $ with a periodic boundary condition. For this purpose it is required to consider the $CFT_{1+1}$ on an infinite cylinder with the spatial direction compactified on a circle of circumference $ L $, as discussed earlier in subsection \ref{subsec_en_dj_int_vac_fin_size}. The corresponding dual bulk configuration in this case is the pure $AdS_3$ space time expressed in global coordinates as follows \cite{Ryu:2006bv,Ryu:2006ef,Cadoni:2009tk,Cadoni:2010kla}
\begin{equation}
\label{eq_metric_pure_ads3_global_coord}
ds^2 = R^2 \left ( - \cosh^2 \rho dt^2 + d\rho^2
+ \sinh^2  \rho d\phi^2 \right ) ,
\end{equation}
where the spatial coordinate $ \phi $ has a  period of $ 2 \pi $. In these coordinates the length 
$ {\cal L}_{\gamma} $ of the bulk space like geodesic anchored on an interval $ \gamma $ (of length $ l_{\gamma} $) may be given as \cite{Ryu:2006bv,Ryu:2006ef,Cadoni:2009tk,Cadoni:2010kla}
\begin{equation}
\label{eq_geo_len_pure_ads3_global_coord}
{\cal L}_{\gamma} = 2 R \ln \left [
\frac{ L }{ \pi a }
\sin \left ( \frac{ \pi l_{\gamma} }{ L } \right )
\right ] ,
\end{equation}
where $ a $ is once again the UV cut of{}f. Utilizing the above expression given in eq.\@ \eqref{eq_geo_len_pure_ads3_global_coord} it is now possible to obtain the holographic entanglement negativity for the mixed state in question from eq.\@ \eqref{eq_holo_ent_neg_dj_int_geo_len} as follows
\begin{equation}
\label{eq_holo_ent_neg_dj_int_finite_size}
{\cal E} = \frac{ 3 R }{ 8 G_N^{(3)} }
\ln \left [
\frac
{
\sin \frac{ \pi \left ( l_1 + l_s \right ) }{ L }
\sin \frac{ \pi \left ( l_2 + l_s \right ) }{ L }
}{
\sin \frac{ \pi l_s }{ L }
\sin \frac{ \pi \left ( l_1 + l_2 + l_s \right ) }{ L }
}
\right ] .
\end{equation}
Note that once again using the Brown-Henneaux formula \cite{Brown:1986nw} we exactly reproduce the $CFT_{1+1}$ replica technique results for the large central charge limit as given in eq.\@ \eqref{eq_ent_neg_dj_int_finite_size}. Interestingly in the adjacent limit we again reproduce the corresponding holographic entanglement negativity for the zero temperature mixed state of adjacent intervals in a finite size system as described in \cite{Jain:2017aqk}.

\subsection{Holographic entanglement negativity for two disjoint intervals at a finite temperature}
\label{subsec_hen_dj_int_fin_temp}

Finally we utilize our conjecture to compute the holographic entanglement negativity for the finite temperature mixed state configuration of disjoint intervals in a $CFT_{1+1}$. In this case the $CFT_{1+1}$ is defined on an infinite cylinder with the Euclidean time direction compactified in a circle of circumference $ \beta \equiv 1 / T $ where $ T $ is the temperature. The corresponding dual bulk $AdS_3$ configuration is now the Euclidean BTZ black hole (black string) at a Hawking temperature $ T $ \cite{Ryu:2006bv,Ryu:2006ef,Cadoni:2009tk,Cadoni:2010kla}. The metric for the Euclidean BTZ black hole is given as
\begin{equation}
\label{eq_metric_ads3_btz_black_hole}
ds^2 = \frac{ \left ( r^2 - r_{h}^2 \right ) }{ R^2 } d\tau^2
+ \frac{ R^2 }{ \left ( r^2 - r_{h}^2 \right ) } dr^2
+ \frac{ r^2 }{ R^2 } d\phi^2 ,
\end{equation}
where $ \tau $ denotes the Euclidean time with the $ \phi $ direction uncompactified and the event horizon is located at $ r = r_h $. For the above bulk $AdS_3$ geometry the corresponding length $ {\cal L}_{\gamma} $ of the bulk space like geodesic anchored on an interval $ \gamma $ (of length $ l_{\gamma} $) is given as follows \cite{Ryu:2006bv,Ryu:2006ef,Cadoni:2009tk,Cadoni:2010kla}
\begin{equation}
\label{eq_geo_len_ads3_btz_black_hole}
{\cal L}_{\gamma} = 2 R \ln \left [
\frac{ \beta }{ \pi a }
\sinh \left ( \frac{ \pi l_{\gamma} }{ \beta } \right )
\right ] ,
\end{equation}
where $ a $ is the UV cut of{}f. As earlier we now utilize the above expression in eq.\@ \eqref{eq_geo_len_ads3_btz_black_hole} to obtain the holographic entanglement negativity for the finite temperature mixed state under consideration from eq.\@ \eqref{eq_holo_ent_neg_dj_int_geo_len} as
\begin{equation}
\label{eq_holo_ent_neg_dj_int_finite_temp}
{\cal E} = \frac{ 3 R }{ 8 G_N^{(3)} }
\ln \left [
\frac
{
\sinh \frac{ \pi \left ( l_1 + l_s \right ) }{ \beta }
\sinh \frac{ \pi \left ( l_2 + l_s \right ) }{ \beta }
}{
\sinh \frac{ \pi l_s }{ \beta }
\sinh \frac{ \pi \left ( l_1 + l_2 + l_s \right ) }{ \beta }
}
\right ] .
\end{equation}
Similar to the previous two cases upon using the Brown-Henneaux formula \cite{Brown:1986nw}, the above expression in eq.\@ \eqref{eq_holo_ent_neg_dj_int_finite_temp} exactly reproduces the corresponding $CFT_{1+1}$ replica results in the large central charge limit given in eq.\@ \eqref{eq_ent_neg_dj_int_finite_temp}. In the adjacent limit our result once again reduces to the corresponding holographic entanglement negativity for the finite temperature mixed state of adjacent intervals described in \cite{Jain:2017aqk}. Naturally the results of the above subsections serve as strong consistency checks for our conjecture which has been obtained through a large central charge analysis for the entanglement negativity of the $CFT_{1+1}$.

\section{Summary and conclusions}
\label{sec_summary}

To summarize we have established a holographic entanglement negativity conjecture involving the bulk geometry for bipartite mixed states of disjoint intervals in a dual $CFT_{1+1}$ through the $AdS_3/CFT_2$ correspondence. In this context we have utilized the large central charge analysis involving the monodromy technique for the entanglement negativity of such mixed states in a holographic $CFT_{1+1}$. Using the large central charge result we have established a holographic construction for the entanglement negativity of the above mixed state configurations, which involves a specific algebraic sum of the lengths of bulk space like geodesics anchored on appropriate intervals. Interestingly the holographic entanglement negativity reduces to an algebraic sum of the holographic mutual informations relevant to a certain combination of the intervals confirming other similar results in the literature.

Application of our conjecture exactly reproduces the entanglement negativity for bipartite mixed states of disjoint intervals in proximity for a holographic $CFT_{1+1}$ obtained through the replica technique, in the large central charge limit and serves as a strong consistency check. In this context we have computed the holographic entanglement negativity for such bipartite mixed states in a $CFT_{1+1}$ for various scenarios. These involve the zero temperature mixed state of disjoint intervals in proximity for both infinite and finite size systems described by a holographic $CFT_{1+1}$. The corresponding bulk dual configurations are described by the pure $AdS_3$ geometry in the Poincar\'{e} and global coordinates respectively. Furthermore we have extended our analysis to obtain the holographic entanglement negativity for the corresponding finite temperature mixed state of such disjoint intervals in a $CFT_{1+1}$ dual to a bulk Euclidean BTZ black hole (black string). Interestingly in each of the scenarios described above we have been able to exactly reproduce the corresponding results for adjacent intervals in a $CFT_{1+1}$ through the adjacent limit which provides further consistency check for our construction.

We would like to mention here that although our holographic entanglement negativity conjecture has been substantiated through applications to specific examples of zero and finite temperature mixed states under consideration, a bulk proof for our conjecture along the lines of \cite{Faulkner:2013yia} is a non trivial open issue that needs attention. Furthermore our analysis suggests a higher dimensional generalization of the holographic entanglement negativity conjecture for such mixed states of disjoint intervals in proximity through the $AdS_{d+1}/CFT_d$ framework. Such an extension would involve a similar algebraic sum of bulk codimension two static minimal surfaces anchored on appropriate subsystems to describe the holographic entanglement negativity for such mixed states under consideration. Naturally such a higher dimensional generalization needs to be substantiated through consistency checks involving applications to specific examples and also a bulk proof along the lines of \cite{Lewkowycz:2013nqa}. Our holographic entanglement negativity conjecture is expected to provide interesting insights into diverse physical phenomena such as topological phases, quantum phase transitions, strongly coupled theories in condensed matter physics and critical issues in quantum gravity, which involve such mixed state entanglement. These constitute fascinating open issues for future investigations.

% ===== ===== ===== Bibliography ===== ===== ===== 

\bibliographystyle{utphys}

\bibliography{hendj_bib}

\end{document}